\documentclass[journal]{IEEEtran}
\usepackage{amsmath,amsfonts}
\usepackage{algorithmic}
\usepackage{algorithm}
\usepackage{array}
\usepackage[caption=false,font=normalsize,labelfont=sf,textfont=sf]{subfig}
\usepackage{textcomp}
\usepackage{stfloats}
\usepackage{url}
\usepackage{verbatim}
\usepackage{graphicx}
\usepackage{cite}
\begin{document}

%ML-assisted Digital Backpropagation with Subband Processing \\ for Nonlinearity Compensation in Coherent Systems
%Subband Learned Digital Backpropagation Framework for Efficient Wideband Nonlinearity Compensation
%ML-assisted Subband Learned Digital Backpropagation for Efficient Wideband Nonlinearity Compensation
\title{ML-assisted Subband Learned Digital Backpropagation \\ for Nonlinearity Compensation in Wideband Optical Systems}

\author{Evgeny Shevelev, Oleg Sidelnikov, Vitaly Danilko, Mikhail Fedoruk, and Alexey Redyuk
        % <-this % stops a space
\thanks{The authors are with Novosibirsk State University, Novosibirsk, 630090, Russia.}% <-this % stops a space
}

\markboth{Journal of \LaTeX\ Class Files,~Vol.~XX, No.~X, May~2026}%
{Shell \MakeLowercase{\textit{et al.}}: A Sample Article Using IEEEtran.cls for IEEE Journals}

\IEEEpubid{}

\maketitle

\begin{abstract}
Digital backpropagation (DBP) is one of the most effective techniques for compensating nonlinear distortions in coherent optical fiber communication systems. However, its practical application to wideband transmission remains limited by high computational complexity caused by large channel memory and the requirement for fine spatial discretization.
In this work, we propose a subband-based learned digital backpropagation (SbL-DBP) framework for wideband optical transmission systems. The received signal is decomposed into multiple subbands, enabling independent frequency-domain compensation of the chromatic dispersion with reduced effective channel memory and lower computational complexity. Nonlinear intra- and inter-subband interactions are addressed in the time domain using a trainable multi-input multi-output filtering structure. The parameters of the proposed framework are jointly optimized using end-to-end gradient-based learning. In addition, sparsification techniques are employed to remove insignificant coefficients and further reduce computational complexity.
Numerical simulations of an 11$\times$40~Gbaud WDM RRC-16QAM 20$\times$100 km transmission system demonstrate that the proposed method provides a superior performance--complexity trade-off compared to conventional DBP and enhanced DBP. In the low- and medium-complexity regimes, SbL-DBP provides higher signal-to-noise ratio  gains while requiring fewer propagation steps.

\end{abstract}


\section{Introduction}
\IEEEPARstart{O}{ptical} fiber communication systems are a key component of modern information infrastructure, supporting global Internet traffic, data center interconnects, and long-haul and transoceanic links. In recent years, the rapid growth of applications and services driven by artificial intelligence has further increased the demand for data transmission, especially within data centers and between them \cite{barnes_2026}. As a result, the need for higher transmission capacity continues to grow \cite{agrell_jo_2024}. This demand is commonly addressed by increasing the symbol rate and expanding the signal bandwidth, leading to wideband transmission systems. However, one of the main limitations in this process is the nonlinear behavior of optical fiber, which distorts the transmitted signal and reduces the achievable spectral efficiency \cite{winzer_oe_2018, soman_jo_2021}. These nonlinear effects, mainly caused by Kerr nonlinearity, become especially important in high-power and wideband systems, effectively turning the transmission link into a nonlinear channel with large memory.

To address this issue, various digital signal processing (DSP) approaches for nonlinearity compensation in optical fiber have been proposed \cite{amari_ieeecst_2017, savory_jstqe_2010, zhao_applsc_2019, redyuk_optcomm_2025}. For practical applications, these approaches must provide high accuracy while maintaining low computational complexity and robustness to noise. Digital backpropagation (DBP) is one of the most well-known methods to compensate for both linear and nonlinear impairments in optical fiber systems. It is based on the numerical solution of the inverse nonlinear Schr\"{o}dinger equation, typically implemented using the split-step Fourier method (SSFM)~\cite{ip_jlt_2008}. Although DBP can provide significant performance gain, its practical use is limited by high computational complexity, since accurate compensation requires a large number of propagation steps involving forward and inverse discrete Fourier transforms (DFT) at each step.
 
To reduce this complexity, several modified DBP approaches have been proposed that retain the implementation of the linear step in the frequency domain while improving overall efficiency \cite{xiao_opex_2017, secondini_ecoc_2014, liang_opex_2015, cellini_arxiv_2026, redyuk_optcont_2025, lin_jlt_2022}. The enhanced split-step Fourier method (ESSFM) introduces a refined nonlinear step that accounts for the interaction between dispersion and nonlinearity, allowing the number of steps to be significantly reduced \cite{secondini_ecoc_2014}. More recent works further extend this idea by optimizing system parameters, such as step lengths and nonlinear filters, including the learned ESSFM framework \cite{cellini_arxiv_2026}. Another important direction is perturbation-based DBP, where nonlinear effects are approximated using an analytical model derived from perturbation theory, leading to reduced computational complexity \cite{redyuk_optcont_2025, lin_jlt_2022}. 

Although these approaches significantly improve the performance--complexity trade-off, they still face important limitations. ESSFM-based methods often rely on simplified nonlinear filters that are identical for all steps and require specialized numerical optimization techniques, which may limit their flexibility in modeling complex signal evolution. Perturbation-based techniques, on the other hand, depend on approximations that may lose accuracy in strongly nonlinear or wideband scenarios. Moreover, most of these methods are designed for full-band processing and do not explicitly address the challenges associated with large channel memory in high symbol rate systems.

In parallel, machine learning (ML) has been widely explored for nonlinear compensation in optical systems \cite{freire_jlt_2024, luo_jlt_2023, sidelnikov_jlt_2021, oliari_jlt_2020, hager_jsac_2021}. In particular, the learned DBP (LDBP) interprets the SSFM structure as a deep neural network and jointly optimizes its parameters using data-driven methods \cite{oliari_jlt_2020, hager_jsac_2021}. 
These approaches can improve the performance--complexity trade-off by adapting the model to channel conditions. However, LDBP and its modifications rely on time-domain compensation of dispersion using finite impulse response (FIR) filters. As a result, the computational complexity scales with channel memory and increases rapidly for high symbol rates and wideband signals, which limits their practical applicability in high-capacity systems.
Moreover, training long sequences of FIR filters is challenging, while finite-length truncation may introduce additional approximation errors.

A key challenge in wideband systems is the large channel memory caused by chromatic dispersion. As the signal bandwidth increases, the dispersion-induced temporal spread of the signal grows, making both frequency-domain and time-domain DBP increasingly demanding in terms of computational complexity due to the requirement for reducing spatial step size. To address this issue, subband processing techniques have been proposed, where the signal spectrum is divided into several narrower bands \cite{fan_jlt_2025, guo_col_2026, hager_ecoc_2018}. Within each subband, the effective channel memory is reduced, enabling more efficient signal processing with shorter filters in both linear and nonlinear steps, as well as fewer required number of spatial steps. However, a non-trivial challenge in this approach is the accurate modeling of inter-subband nonlinear effects. Recent methods, such as subband DBP \cite{ip_ofc_2011} and coupled-band ESSFM \cite{civelli_jlt_2025}, improve performance by accounting for interactions between subbands and modeling cross-phase modulation effects more accurately. Nevertheless, these methods typically rely on predefined or analytically derived models of inter-subband interaction, which may limit their flexibility in complex transmission scenarios.

In this work, we propose a novel low-complexity DBP framework based on subband processing, which integrates ML-assisted joint optimization of nonlinear phase rotation filters within a DFT-based SSFM structure, enabling flexible and accurate modeling of both intra- and inter-subband nonlinear interactions. To address the large channel memory in wideband systems, the received signal is decomposed into multiple subbands, where linear effects are compensated independently in each subband in the frequency domain using DFT-based operations. In the nonlinear step, both intra-subband and inter-subband interactions are modeled in the time domain. Within each subband, nonlinear phase rotation is computed using convolution of neighboring samples with a trainable filter, capturing intra-subband effects. Inter-subband nonlinear interactions are modeled through a set of coupled filtering operations, where the phase correction in each subband depends on the signal power in neighboring subbands. These interactions are implemented as a multi-input multi-output (MIMO) filtering structure with memory, mapping subband intensities to phase rotations through a set of learnable coefficients. All model parameters are jointly optimized using end-to-end learning. This data-driven optimization enables the model to adapt to channel conditions without relying on predefined analytical approximations. Furthermore, sparsification techniques are applied during training to remove insignificant coefficients, thereby reducing computational cost while preserving performance.

\section{Methods}

\subsection{System Model}
We consider a coherent optical communication system with single-polarization transmission, where signal propagation in the fiber is governed by chromatic dispersion, Kerr nonlinearity, and attenuation.
The evolution of the optical signal can be described by the nonlinear Schr\"{o}dinger equation (NLSE):
\begin{equation}
\frac{\partial A}{\partial z} = \left(-\frac{\alpha}{2} - i\frac{\beta_2}{2}\frac{\partial^2}{\partial t^2} + i\gamma|A|^2\right) A,
\label{eq:nlse}
\end{equation}
where $A(z,t)$ is the complex envelope of the optical field, $z$ is the propagation distance, and $t$ is time. The parameters $\alpha$, $\beta_2$ and $\gamma$ denote the fiber attenuation, group-velocity dispersion, and the Kerr nonlinearity coefficient, respectively.
The extension of the proposed approach to dual-polarization systems is straightforward.

At the transmitter, the signal is generated as a sequence of modulated symbols shaped by a pulse shaping filter:
\begin{equation}
A(z=0, t) = \sqrt{P_0}\sum_{k} a[k] f(t - kT), 
\end{equation}
where $P_0$ is the launch power, $a[k]$ are complex data symbols, $f(t)$ is the pulse shape, and $T$ is the symbol interval.

In practical long-haul systems, periodic optical amplification is used to compensate for fiber losses. In this work, we consider lumped amplification using erbium-doped fiber amplifiers (EDFAs), which introduce amplified spontaneous emission (ASE) noise. As a result, the received signal is affected by both deterministic impairments described by Eq.~(\ref{eq:nlse}) and additive noise.

Signal propagation in optical fiber links is typically modeled using SSFM. 
In this approach, the fiber link is divided into short segments and signal evolution over each segment is approximated by alternating linear and nonlinear operations in accordance with Eq.~(\ref{eq:nlse}). The linear step accounts for dispersion and attenuation in the frequency domain, while the nonlinear step models the Kerr effect in the time domain. The accuracy of this approximation depends on the number of segments used.

In the following, we focus on wideband wavelength-division multiplexing (WDM) transmission systems, where a large signal bandwidth leads to significant temporal spreading due to chromatic dispersion, resulting in increased channel memory. This effect makes nonlinear compensation more challenging and motivates the development of efficient DBP-based methods.

\subsection{Digital Backpropagation}
\label{subsec:dbp}
Digital backpropagation is a widely studied multi-step technique for compensating deterministic signal distortions in optical fiber systems, including both linear and nonlinear impairments \cite{ip_jlt_2008}. The method is based on the numerical solution of the inverse NLSE, where the received signal is digitally propagated backward along the transmission link. This is achieved by applying the SSFM with reversed fiber parameters, allowing for reconstruction of the transmitted signal.

In DBP, the fiber parameters are inverted as $\alpha'=-\alpha$, $\beta_2'=-\beta_2$, and $\gamma'=-\gamma$. Inverse propagation is approximated using SSFM, where each step consists of a linear and a nonlinear operation applied to the discrete-time signal. Let $\mathbf{u}^{(k)} = [u^{(k)}[n]]$ denote the signal at the $k$-th step.
A single step of the symmetric SSFM can be expressed as follows.
The linear operator is implemented in the frequency domain as
\begin{equation}
\mathbf{u}_\mathrm{L} = \mathcal{F}^{-1} \left\{ H_\mathrm{L}(\omega, \Delta z_1) \cdot \mathcal{F}\{\mathbf{u}^{(k)}\} \right\},
\end{equation}
where $\mathcal{F}\{\cdot\}$ and $\mathcal{F}^{-1}\{\cdot\}$ denote DFT and inverse DFT, respectively, and 
\begin{equation}
H_\mathrm{L}(\omega, \Delta z) = \exp\left[\left(-\frac{\alpha'}{2} + i\frac{\beta_2'}{2}\omega^2\right)\Delta z \right]
\end{equation}
is the linear transfer function describing dispersion and attenuation.
The nonlinear operator is applied in the time domain as
\begin{equation}
u_\mathrm{NL}[n] = u_\mathrm{L}[n] \cdot \exp\left(i\gamma' |u_\mathrm{L}[n]|^2 \Delta z \right),
\label{ssfm_nln}
\end{equation}
which accounts for Kerr nonlinearity through a phase rotation proportional to the instantaneous signal power.
Combining both operations, a symmetric SSFM step can be written as
\begin{equation}
\mathbf{u}^{(k+1)} = \mathcal{L}_{\Delta z_2} \left( \mathcal{N}_{\Delta z} \left( \mathcal{L}_{\Delta z_1} (\mathbf{u}^{(k)}) \right) \right),
\end{equation}
where $\mathcal{L}_{\Delta z}(\cdot)$ and $\mathcal{N}_{\Delta z}(\cdot)$ denote linear and nonlinear operators, respectively, and $\Delta z_1 = \lambda \Delta z$, $\Delta z_2 = (1-\lambda)\Delta z$.
For $\lambda = 0.5$, the conventional symmetric SSFM scheme is obtained~\cite{agrawal_2013}, which provides second-order approximation accuracy with respect to the step size $\Delta z$. However, this choice is not necessarily optimal in the context of DBP. When $\lambda \neq 0.5$, the scheme formally reduces to first-order accuracy, but an appropriate selection of $\lambda$ may still improve nonlinear compensation performance in practice \cite{civelli_jlt_2025}.
Throughout this work, the parameter $\lambda$ was selected empirically based on the overall system performance.

By cascading multiple steps, the full inverse propagation over the fiber link is approximated.
In this formulation, the internal linear steps are applied with a step size of $\Delta z$, while the first and last linear operations are performed with step sizes of $\lambda \Delta z$ and $(1-\lambda) \Delta z$, respectively.

It is well known that conventional DBP is computationally demanding due to the repeated application of forward and inverse Fourier transforms, especially when small step sizes are required to ensure high accuracy. To reduce this complexity, several modified DBP approaches have been proposed, including ESSFM \cite{secondini_ecoc_2014}, which modifies the nonlinear step to account for the interaction between dispersion and nonlinearity within each step. In conventional DBP, the nonlinear phase rotation depends only on the instantaneous signal power, as shown in Eq.~(\ref{ssfm_nln}).
In contrast, ESSFM introduces a memory effect by incorporating neighboring samples:
\begin{equation}
u_\mathrm{NL}[n] = u_\mathrm{L}[n]\cdot\exp\left(i\gamma'\!\! \sum_{k=-N_c}^{N_c} \!\!C_k\, |u_\mathrm{L}[n+k]|^2 \Delta z \right),
\end{equation}
where $N_c$ defines the number of neighboring samples considered, and $C_k$ are real-valued coefficients that model the interplay between dispersion and nonlinearity. When $N_c = 0$ and $C_0 = 1$, the expression reduces to the nonlinear step of conventional DBP. 
By incorporating this extended nonlinear interaction, ESSFM can significantly reduce the required number of steps while maintaining high accuracy. However, in most implementations, the coefficients $C_k$ are predefined or shared across steps, which may limit the flexibility of the model in complex transmission scenarios.

While these approaches improve the performance--complexity trade-off, they remain limited by fixed model structures and do not explicitly address the challenges associated with wideband signals.
In particular, due to chromatic dispersion, effective channel memory increases quadratically with signal bandwidth~\cite{savory_oe_2008}, leading to a significant growth in the temporal span of nonlinear interactions. As a result, accurate compensation requires long nonlinear filters and small spatial step sizes, which in turn leads to high computational resources. 
Subband processing provides a natural way to mitigate these limitations.
By decomposing the signal into multiple subbands, the bandwidth of each component is reduced, which shortens the corresponding channel memory and enables more efficient processing.
In addition, the reduced bandwidth and lower signal power within each subband relax the constraints on the SSFM step size. Specifically, narrower bandwidth reduces dispersion-induced temporal variations, while lower power decreases nonlinear phase accumulation, allowing larger spatial step sizes $\Delta z$ to be used while maintaining sufficient numerical accuracy.
Motivated by these observations, we propose a subband-based DBP framework with learnable nonlinear interactions, enabling flexible and efficient modeling of both intra- and inter-subband effects.

\subsection{Proposed Subband Learned DBP Framework}
\begin{figure*}[!t]
\centering
\includegraphics[width=7in]{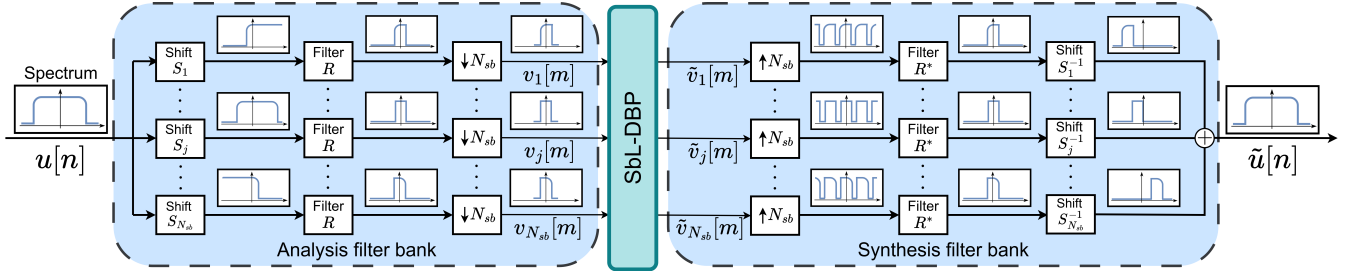}
\caption{Scheme of a proposed NLC consisting of analysis filter bank, a subband learned DBP module and synthesis filter bank with $N_{sb}$ subbands.}
\label{full_scheme}
\end{figure*}

The proposed nonlinear compensation (NLC) framework is based on subband processing, which enables efficient handling of wideband signals with large channel memory~\cite{hager_ecoc_2018}. The general structure of NLC is illustrated in Fig.~\ref{full_scheme}. The method combines subband decomposition with a learned DBP algorithm operating in each subband, followed by signal reconstruction. The framework consists of three key blocks: an analysis filter bank (AFB), a subband learned DBP (SbL-DBP) module, and a synthesis filter bank (SFB). In the following subsections, we describe each part of the proposed approach, including signal decomposition, linear and nonlinear processing in the subband domain, and reconstruction of the full-band signal.

\subsubsection{Subband decomposition and signal representation}
\begin{figure}[htbp]
\centering
\includegraphics[width=3.4in]{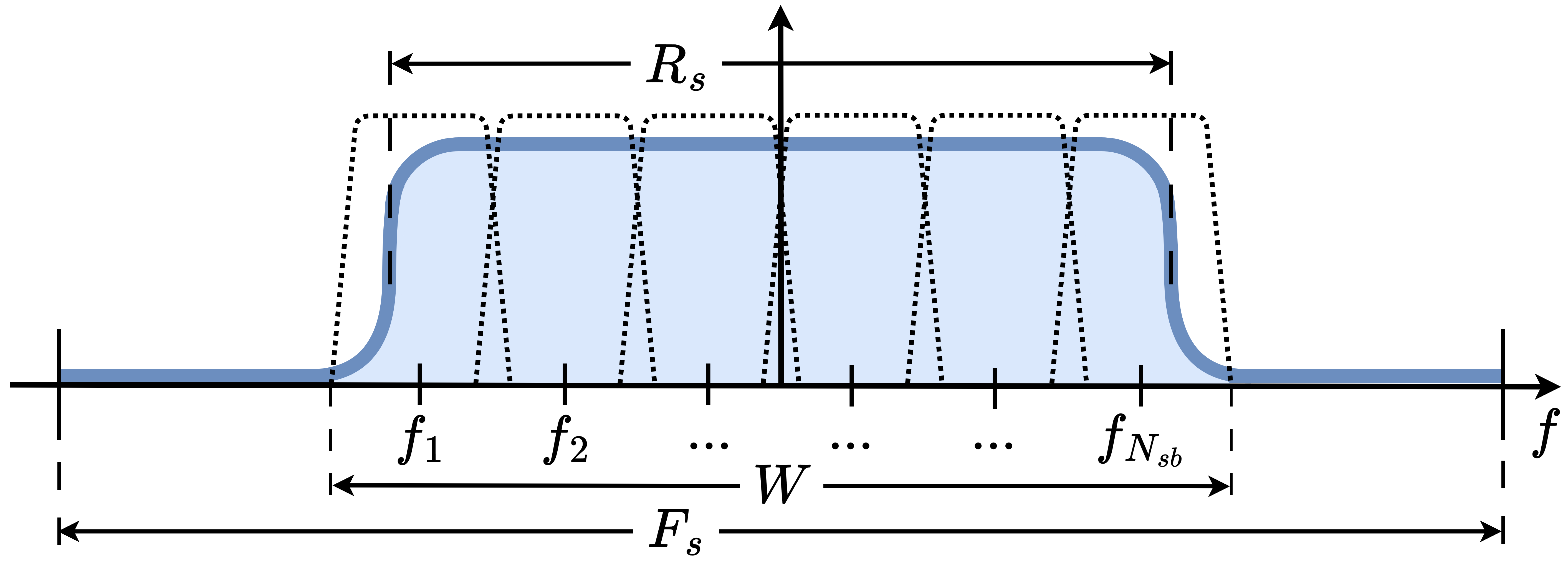}
\caption{Scheme of the wideband signal spectrum: $W$ -- spectral support of the useful signal components, $F_s$ -- sampling rate, $R_s$ -- symbol rate, $f_j$ -- center frequency of the $j$-th subband, $j \in \{1, \dots, N_{sb}\}$.}
\label{spec}
\end{figure}

The analysis filter bank performs the decomposition of the wideband input signal $u[n]$ into $N_{sb}$ subbands. 
This operation is implemented through a sequence of frequency shifting, filtering, and downsampling steps, as illustrated in Fig.~\ref{full_scheme}.
The input to the analysis filter bank is a wideband signal sampled at a rate $F_s$, typically satisfying $F_s \geq R_s$, where $R_s = 1/T$ is the symbol rate. 
Let $W$ denote the spectral support of the useful signal components, as schematically illustrated in Fig.~\ref{spec}.
The center frequency of the $j$-th subband is defined as $f_j = -\frac{W}{2} + \frac{W}{N_{sb}}\left(j - \frac{1}{2}\right), j \in \mathcal{J} = \{1, \dots, N_{sb}\}$.
Following a standard filter bank implementation, the analysis stage applies a frequency shift operator $S_j(\cdot)$ to each subband.
This operator multiplies the input discrete-time signal $u[n]$ by a complex exponential $S_j\left(u[n]\right)=u[n]e^{-i2\pi f_j n/F_s}$,
which shifts the signal spectrum by $-f_j$, aligning the center frequency of the $j$-th subband with zero frequency.
After the frequency shift, a band-pass filter $R(\cdot)$ is applied to remove out-of-band components and isolate the desired subband.
The resulting signal is then downsampled by a factor of $N_{sb}$, reducing the sampling rate proportionally to the bandwidth of each subband.
As a result, the original wideband signal $u[n]$ is represented as a set of $N_{sb}$ subband signals $v_j[m]$ with reduced bandwidth and sampling rate.
This representation reduces the effective channel memory in each subband and enables more efficient subsequent processing.

\subsubsection{Subband processing}
\begin{figure}[!htbp]
\centering
\includegraphics[width=3.4in]{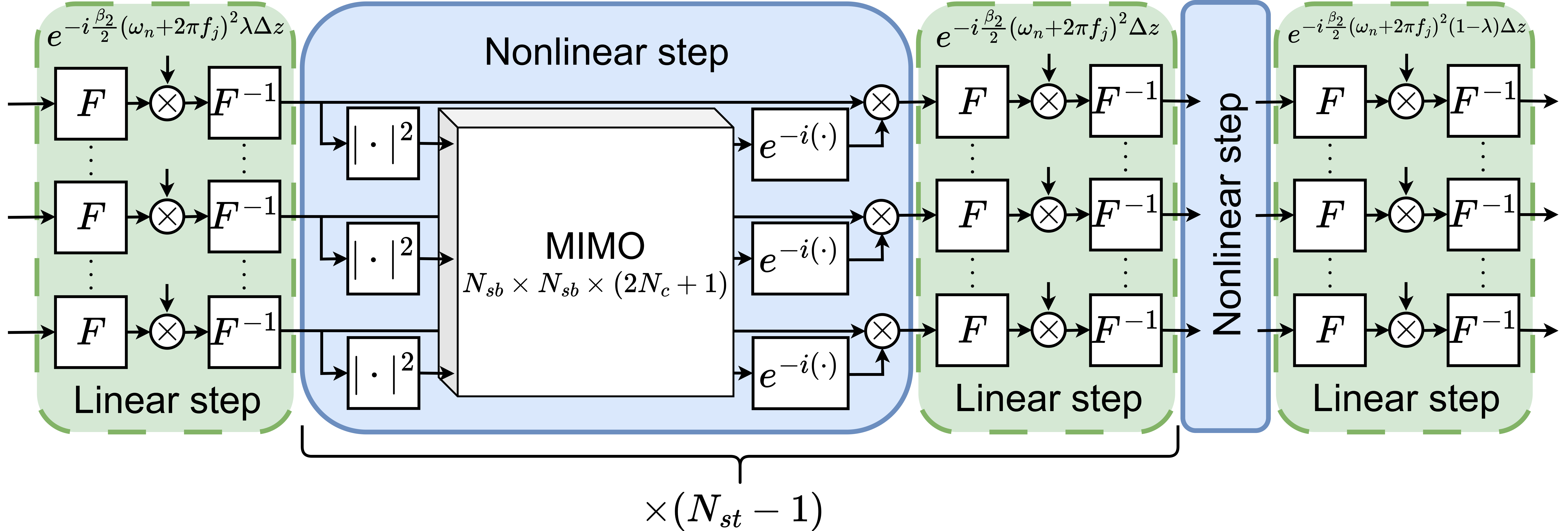}
\caption{Structure of the proposed SbL-DBP framework with $N_{st}$ steps. Linear steps (green blocks) are performed independently in each subband in the frequency domain, nonlinear steps (blue blocks) model intra- and inter-subband interactions in the time domain using MIMO filtering.}
\label{DBP_scheme}
\end{figure}

The set of subband signals $v_j[m]$, obtained after AFB, is fed into the SbL-DBP processing block.
The overall processing flow is illustrated in Fig.~\ref{DBP_scheme} and summarized in Algorithm~\ref{alg1}.
The proposed approach follows the structure of symmetric DBP, introduced in Section~\ref{subsec:dbp}, while extending it to the subband domain.
In the proposed framework, each DBP step consists of a linear and a nonlinear operation.
The linear step is performed independently in each subband in the frequency domain, while the nonlinear step is carried out in the time domain and accounts for both intra-subband and inter-subband interactions.
In the linear step, chromatic dispersion is compensated separately for each subband using DFT-based processing. Since the subband signals are represented in the baseband domain after analysis filtering, the original spectral position of each subband is restored in the linear operator by replacing $\omega$ with $\omega + 2\pi f_j$, as shown in Fig.~\ref{DBP_scheme} and Algorithm~\ref{alg1}. 
In the nonlinear step, the signal evolution is modeled using an extended nonlinear phase rotation that incorporates both intra-subband and inter-subband effects.
Specifically, the nonlinear update for the $j$-th subband is given by
\begin{equation}
    w_j[m] = v_j[m] \cdot \exp\Big(i \gamma' \eta_s\sum_{l\in \mathcal{J}} \sum_{k=-N_c}^{N_{c}}\!\! C_{jlk}\big|v_l[m+k]\big|^2  \Delta z\Big),
    \label{eq:sbl_dbp}
\end{equation}
where $j$ denotes the index of the current subband, $l$ indexes interacting subbands, and $m$ is the discrete-time sample index. Similarly to ESSFM, the parameter $N_c$ defines the memory of the nonlinear interaction, i.e., the number of neighboring samples taken into account. The index $s = 1, \dots, N_{st}$ denotes the propagation step, where $N_{st}$ is the total number of steps, and $\eta_s$ is a step-dependent scaling factor.
This formulation generalizes the conventional DBP nonlinear step by incorporating both temporal and inter-subband dependencies.

The coefficients $C_{jlk}$ are real-valued parameters that determine the contribution of the $l$-th subband at time offset $k$ to the nonlinear phase rotation in the $j$-th subband.
For $l = j$, the coefficients $C_{jjk}$ describe intra-subband nonlinear effects, specifically self-phase modulation (SPM) and its interaction with dispersion.
For $l \neq j$, the coefficients $C_{jlk}$ capture inter-subband interactions, such as cross-phase modulation (XPM) between different spectral components.
It should be noted that the proposed formulation does not explicitly account for four-wave mixing (FWM) interactions between subbands.
In this way, the nonlinear step can be interpreted as a MIMO filtering operation, where the phase rotation in each subband depends on the intensity profiles of all subbands within a finite temporal window.
When $N_{sb} = 1$, $N_c = 0$ and $C_{110}=1$, the expression in (\ref{eq:sbl_dbp}) reduces to the standard nonlinear step of conventional DBP.
The case with $N_{sb} = 1$, $N_c > 0$ is referred to as enhanced DBP (EnDBP).
The coefficients $C_{jlk}$ are treated as trainable parameters and optimized using the end-to-end learning framework described in Section~\ref{subsec:training}.

    \begin{algorithm}
    \caption{SbL-DBP processing for the $j$-th subband}
    \begin{algorithmic}[H]
    \STATE \textbf{Input:} subband signals $\{ \mathbf{v_j} \}_{j \in \mathcal{J}}$ from AFB
    \STATE $\mathbf{\hat v_j} \gets \mathcal{F}\left(\mathbf{v_j}\right)$
    \STATE $\hat v_j[m] \gets \hat v_j[m] \cdot e^{-i \beta_2 \left(\omega_m + 2\pi f_j\right)^2 \lambda \Delta z / 2}$
    \STATE $\mathbf{v_j} \gets \mathcal{F}^{-1}\left(\mathbf{\hat v_j}\right)$
    % \STATE $w_j[k] = v_j[k] \cdot \exp\left(-i \eta_1 \gamma \big|v_j[k]\big|^2 \Delta z\right)$
    \STATE $w_j[m] \gets v_j[m] \cdot e^{-i \gamma \eta_1\sum_{l\in \mathcal{J}} \sum_{k=-N_c}^{N_{c}} C_{jlk}\big|v_l[m+k]\big|^2  \Delta z}$
    \FOR{$s \in [2,N_{st}]$}
        \STATE $\mathbf{\hat v_j} \gets \mathcal{F}\left(\mathbf{w_j}\right)$
        \STATE $\hat v_j[m] \gets \hat v_j[m] \cdot e^{-i \beta_2 \left(\omega_m + 2\pi f_j\right)^2\Delta z / 2 }$
        \STATE $\mathbf{v_j} \gets \mathcal{F}^{-1}\left(\mathbf{\hat v_j}\right)$
        % \STATE $w_j[k] = v_j[k] \cdot \exp\left(-i \eta_s \gamma \big|v_j[k]\big|^2 \Delta z\right)$
        \STATE $w_j[m] \gets v_j[m] \cdot e^{-i \gamma \eta_s\sum_{l\in \mathcal{J}} \sum_{k=-N_c}^{N_{c}} C_{jlk}\big|v_l[m+k]\big|^2  \Delta z}$
    \ENDFOR
    \STATE $\mathbf{\hat v_j} \gets \mathcal{F}\left(\mathbf{w_j}\right)$
    \STATE $\hat v_j[m] \gets \hat v_j[m] \cdot e^{-i \beta_2 \left(\omega_m + 2\pi f_j\right)^2 (1-\lambda) \Delta z / 2 }$
    \STATE $\mathbf{\tilde{v}_j} \gets \mathcal{F}^{-1}\left(\mathbf{\hat v_j}\right)$
    \STATE \textbf{Output:} processed subbands $\{ \mathbf{\tilde{v}_j} \}_{j \in \mathcal{J}}$ for SFB
    \end{algorithmic}
    \label{alg1}
    \end{algorithm}

\subsubsection{Subband synthesis and reconstruction}
The synthesis filter bank performs the reconstruction of the wideband signal $\tilde{u}[n]$ from the set of $N_{sb}$ processed subband signals $\tilde{v}_j[m]$.
This operation is carried out by applying a sequence of steps that are inverse to those used in AFB, as shown in Fig.~\ref{full_scheme}.
In the first stage, each subband signal is upsampled to the original sampling rate. This is achieved by inserting $N_{sb}-1$ zeros between consecutive samples, thereby increasing the sampling rate by a factor of $N_{sb}$.
After upsampling, a reconstruction filter $R^*(\cdot)$ is applied to each subband signal. This filtering step suppresses the spectral images introduced by the upsampling process and restores the desired spectral shape of the signal.
Next, an inverse frequency shift operator $S_j^{-1}(\cdot)$ is applied to each subband, translating it from the baseband back to its original center frequency $f_j$.
Finally, all reconstructed subband signals are summed to obtain the full-band signal.
With properly designed analysis and synthesis filters, this process ensures an accurate (ideally perfect) reconstruction of the original signal.
After SFB, the reconstructed signal is passed through a matched filter and the sampling rate is reduced to one sample per symbol for subsequent demodulation.

\subsection{End-to-end Training of SbL-DBP Parameters}
\label{subsec:training}
\begin{figure}[!htbp]
\includegraphics[width=3.4in]{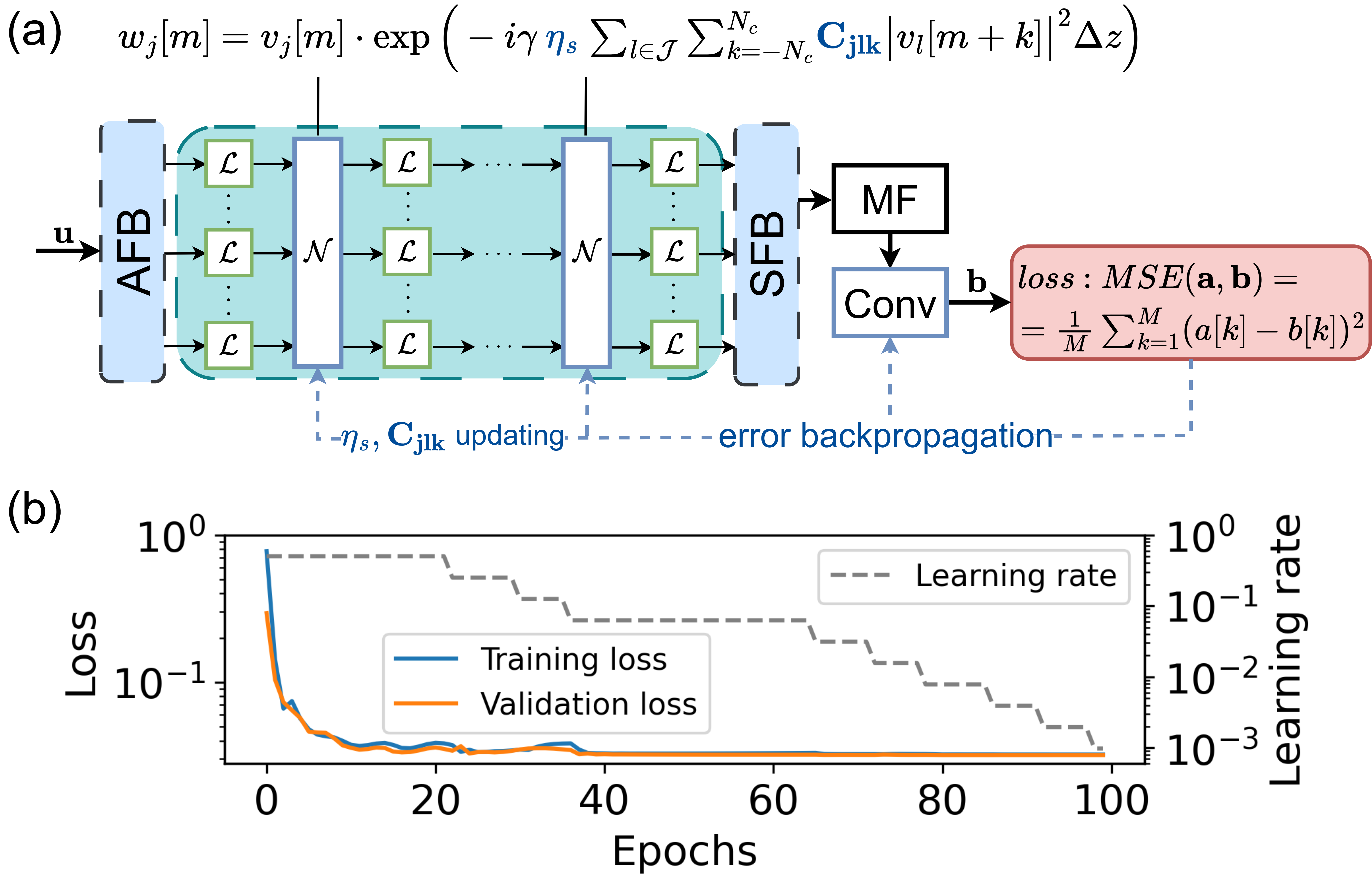}
\caption{(a) End-to-end learning framework of the SbL-DBP, represented as a deep layered model with alternating linear and nonlinear blocks. The MIMO coefficients \( C_{jlk} \) are optimized jointly using gradient-based training. (b)~Dynamics of training loss, validation loss, and learning rate, as functions of the training epoch.}
\label{DBP_learn_scheme}
\end{figure}

The proposed SbL-DBP framework introduces a set of trainable parameters in the nonlinear processing blocks, which enables flexible modeling of both intra- and inter-subband interactions. To efficiently determine these parameters, an end-to-end learning approach is employed, where the entire processing chain is optimized jointly. 
The SbL-DBP algorithm is implemented using framework PyTorch and is represented as a deep layered model composed of alternating linear and nonlinear blocks, following the structure of the SSFM, as illustrated in Fig.~\ref{DBP_learn_scheme}(a). In this representation, each SbL-DBP step corresponds to a layer of the model, which allows the use of gradient-based optimization via error backpropagation for joint training of all model parameters.

The trainable parameters of the model include the scaling coefficients $\eta_s$, the MIMO coupling coefficients $C_{jlk}$, and the coefficients of the convolution-based decimation filter (Conv block in Fig.~\ref{DBP_learn_scheme}(a)). After subband recombination by the SFB and matched filtering, the signal is downsampled to one sample per symbol using this trainable filter. Joint optimization of the decimation filter together with the nonlinear SbL-DBP parameters enables adaptive compensation of residual distortions prior to symbol detection and improves the overall end-to-end performance of the receiver.
The coefficients $\eta_s$ and $C_{jlk}$ are initialized to zero, corresponding to an initial model without nonlinear compensation beyond the linear SbL-DBP structure. This neutral initialization avoids imposing prior assumptions on the interaction model and allows the training process to adaptively learn the relevant nonlinear dependencies from the data.

The training is performed in a supervised manner by minimizing the loss function as the mean squared error (MSE) between the transmitted symbols $\mathbf{a}$ and the corresponding equalized output $\mathbf{b}$:
\begin{equation}
L = \frac{1}{M} \sum_{k=1}^{M} \left| a[k] - b[k] \right|^2.
\end{equation}
This loss function directly reflects the quality of the signal reconstruction. Parameter updates are carried out using the Adam optimizer.

To reduce computational complexity, sparsification is applied to the learned coefficients. 
In particular, a $\ell_1$-regularization term is added to the loss function to promote sparsity by penalizing small-magnitude coefficients:
\begin{equation}
L_{\text{total}} = L + \theta \left(\sum_{s} |\eta_{s}| + \sum_{j,l,k} |C_{jlk}| \right),
\label{eq:reg}
\end{equation}
where $\theta$ controls the strength of the regularization.
This encourages insignificant coefficients to approach zero during training, enabling their removal through pruning.
As a result, the nonlinear operator acquires a sparse structure that reduces computational complexity while preserving the dominant nonlinear interactions.
This also improves interpretability by identifying the most relevant inter-subband interactions.

To ensure robust generalization performance, the dataset is split into three independent subsets: training, validation, and test sets. Each subset consists of sequences of length $2^{19}$ samples.
The training set is used to optimize the model parameters, the validation set is used to adjust hyperparameters and monitor convergence, and the test set is used only for the final evaluation of the trained model.

During training, both training and validation sets are further segmented into blocks of size $2^{12}$ samples, which are organized into minibatches of three blocks each. The optimization is initialized with a learning rate of $0.5$. A learning rate scheduler based on validation performance is applied, reducing the learning rate on plateau with a patience of 5 epochs and a reduction factor of $0.5$.

Fig.~\ref{DBP_learn_scheme}(b) shows the typical example of training and validation loss curves, along with the evolution of the learning rate. Both losses decrease rapidly during the initial training phase, indicating efficient convergence. After approximately 30--40 epochs, the losses stabilize and reach a plateau, suggesting that the model has approached its optimal performance for the given data. The small gap between training and validation loss throughout training indicates good generalization and absence of significant overfitting. The learning rate is reduced stepwise according to the plateau-based scheduler, further supporting stable convergence.

\subsection{Computational Complexity Analysis}
In this section, we analyze the computational complexity of the proposed SbL-DBP method in terms of the number of real multiplications per processed symbol (RMpS), taking into account both linear and nonlinear operations.
We consider a signal block $u[n]$ consisting of $N$ samples, which is decomposed into $N_{sb}$ subbands. Each subband signal $v_j[m]$ contains $N'=N/N_{sb}$ samples.
The total number of DBP steps is indicated by $N_{st}$, and $N_d$ symbols are discarded to mitigate edge effects introduced by block processing.

\subsubsection{Linear step}
In the linear step, the fast Fourier transform (FFT) is used to compensate for chromatic dispersion. A single FFT (or inverse FFT) of length $M$ requires approximately $\frac{M}{2}\log_2 M$ complex multiplications. 
In the proposed approach, the signal is processed in $N_{sb}$ subbands, each of length $N'$.
Therefore, the total number of complex multiplications for the single linear step is given by:
\begin{equation}
\left(N'\log_2N' + N'\right) N_{sb} = \left(N\log_2 N + N - N\log_2 N_{sb}\right).
\end{equation}
This expression shows that subband processing reduces the FFT complexity by $N\log_2N_{sb}$.
Taking into account all linear operations within the SSFM structure, including the initial and final steps, the total complexity of the linear part per symbol, expressed in terms of real multiplications, is given by:
\begin{equation}
C_l = 4 (N_{st} + 1) \frac{N\log_2 N + N - N\log_2 N_{sb}}{(R_s/F_s)N - N_d},
\label{eq:cln}
\end{equation}
where factor 4 accounts for the conversion from complex to real multiplications, and the term $(R_s/F_s)N$ represents the number of transmitted symbols corresponding to $N$ processed samples.

\begin{figure*}[!t]
\centering
\includegraphics[width=7in]{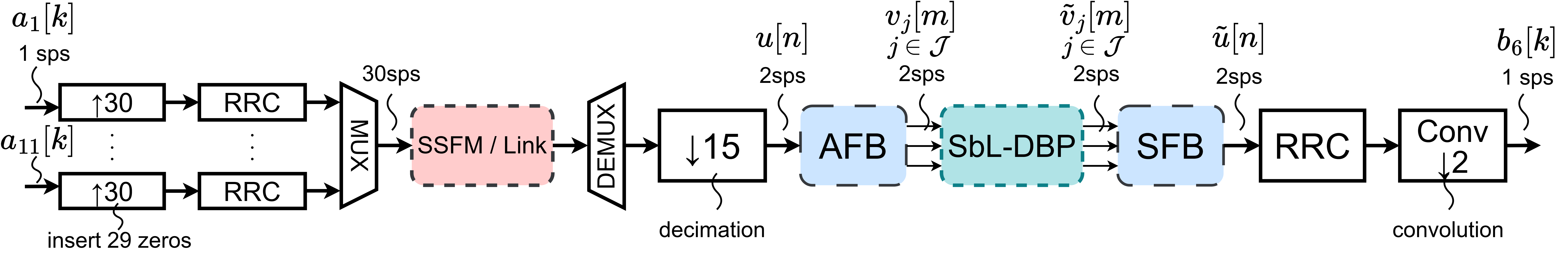}
\caption{Simulation and DSP pipeline of the considered WDM system. RRC -- root-raised cosine, MUX -- multiplexer, SSFM -- split-step Fourier method, DEMUX -- demultiplexer, AFB -- analysis filter bank, SbL-DBP -- subband learned digital backpropagation, SFB -- synthesis filter bank, Conv -- convolution.}
\label{pipeline}
\end{figure*}

\subsubsection{Nonlinear step}
The nonlinear step is implemented in the time domain using a memory-based MIMO filtering structure.
Its complexity is determined by several components.
First, computing the squared magnitudes of the complex samples $|v_j[m]|^2$ requires two real multiplications per sample. Since the magnitudes are computed once for all samples in all subbands, this operation requires a total of $2N' N_{sb}$ real multiplications. 
Second, the weighted summation with the MIMO coefficients $C_{jlk}$ requires $N_{sb} N_{sb} (2N_c + 1)$ real multiplications. 
Third, multiplication by the scalar factor $\gamma \Delta z$ requires $N_{sb}$ real multiplications. 
Finally, applying the complex phase rotation for each subband requires four real multiplications per sample.
Assuming that the exponential function is implemented via a lookup table, its computational cost is neglected. 
Under these assumptions, the total complexity of the nonlinear step per symbol is given by:
\begin{equation}
\begin{aligned}
C_{nl} &= N_{st} \frac{N' N_{sb}\left(2 + N_{sb} (2N_c + 1) + 1 + 4\right)}{(R_s/F_s)N - N_d} \\
&= N_{st} \frac{N \left(N_{sb}\left(2N_c + 1\right) + 7\right)}{(R_s/F_s)N - N_d}.
\end{aligned}
\label{eq:cnln_MIMO}
\end{equation}

\subsubsection{Total complexity}
The total computational complexity of the proposed SbL-DBP method is given by the sum of linear and nonlinear contributions: $C_{\text{total}} = C_l + C_{nl}$.
The obtained expressions highlight important trade-offs of the proposed approach.
While subband decomposition reduces the complexity of the linear step by decreasing the FFT size, the nonlinear step introduces additional computations due to inter-subband interactions.
However, the use of sparsification techniques, as described in Section~\ref{subsec:training}, significantly reduces the effective number of MIMO coefficients, leading to a favorable balance between accuracy and computational complexity.
A detailed investigation of the sparsification procedure and the resulting complexity reduction is presented in Section~\ref{subsec:mimo-coeff}.
It should be mentioned that this analysis does not account for the effects of memory access and subband parallelization, which may further improve the practical efficiency of the proposed method.

\section{Results and Discussion}
In this section, we evaluate the proposed SbL-DBP method in terms of performance, complexity, and interpretability. We first analyze the impact of subband decomposition and the number of propagation steps, then study the structure of the learned nonlinear interactions, and finally investigate the trade-offs between performance and computational complexity.

\subsection{Simulation Setup}
We evaluate the proposed method using numerical simulations of a wideband WDM transmission system. Fig.~\ref{pipeline} shows the overall simulation and post-processing scheme. The transmitted signal consists of 11 WDM channels with a symbol rate of 40~GBaud and 16-QAM modulation. The channel spacing is set to 50~GHz and the launch power is fixed at 1~dBm per channel. Each channel is pulse-shaped using a root-raised cosine (RRC) filter with a roll-off factor of 0.1.

The transmission link comprises 20 fiber spans, each consisting of 100~km of standard single-mode fiber. The fiber parameters are attenuation $\alpha = 0.2$~dB/km, group velocity dispersion $\beta_2 = -21.7$~ps$^2$/km, and nonlinear coefficient $\gamma = 1.2$~W$^{-1}$km$^{-1}$. After each span, EDFAs with a noise figure of 4.5~dB are used to compensate for fiber loss. The amplifiers introduce ASE noise, which is modeled as additive white Gaussian noise and accumulates along the transmission link. Each simulation run uses $2^{18}$ transmitted symbols per WDM channel. Multiple independent realizations with randomly generated symbol sequences and ASE noise are used to construct the training, validation, and test datasets.

The propagation of the signal along the link is governed by the NLSE~\eqref{eq:nlse} and is solved numerically using the symmetric SSFM. In the fiber simulation, a sampling rate of 30 samples per symbol (sps) is used. The step size within each span follows a logarithmic distribution with a finer resolution near the span input, where the signal power is higher. Specifically, the step size is selected such that the nonlinear phase shift accumulated over each step does not exceed $10^{-3}$ rad.

At the receiver, the central WDM channel is selected using an ideal rectangular bandpass filter and then downsampled from 30 to 2 sps, corresponding to a decimation factor of 15. The resulting signal is processed by the proposed NLC framework, which consists of AFB, SbL-DBP and SFB blocks, as shown in Fig.~\ref{full_scheme}. 
For this stage we used $R_s=40$~GBaud, $W=44$~GHz, $F_s=80$~GHz, $N=32768$, and $N_d=256$.
The analysis and synthesis filters $R(\cdot)$ and $R^*(\cdot)$ are implemented as rectangular frequency-domain filters with bandwidth $W/N_{sb}$ and no overlap between neighboring subbands.

After nonlinear compensation, a matched RRC filter is applied, followed by downsampling to one sample per symbol using a convolution-based filter. Symbol detection is then performed using a standard 16-QAM demodulator.

The system performance is evaluated in terms of signal-to-noise ratio (SNR) at the receiver, computed from the measured bit error rate (BER) as
\begin{equation}
    SNR = 20 \log_{10}\left[ \sqrt{10} \, \mathrm{erfc}^{-1}\left( 8 BER / 3 \right)\right],
\end{equation}
where $\mathrm{erfc}^{-1}(\cdot)$ denotes the inverse complementary error function.
In addition, performance improvement is quantified as $\Delta SNR = SNR_{NLC} - SNR_{CDC}$, where $SNR_{CDC}$ corresponds to the SNR after compensation of chromatic dispersion only, and $SNR_{NLC}$ is the SNR after applying the proposed NLC method. For the considered link configuration and a launch power of 1~dBm, the baseline performance is $BER_{CDC}= 0.0144$ and $SNR_{CDC}= 13.3$~dB.

\subsection{Impact of Subband Decomposition and Number of Steps}
\label{subsec:subband_and_num_steps}
\begin{figure}[!htbp]
\includegraphics[width=3.4in]{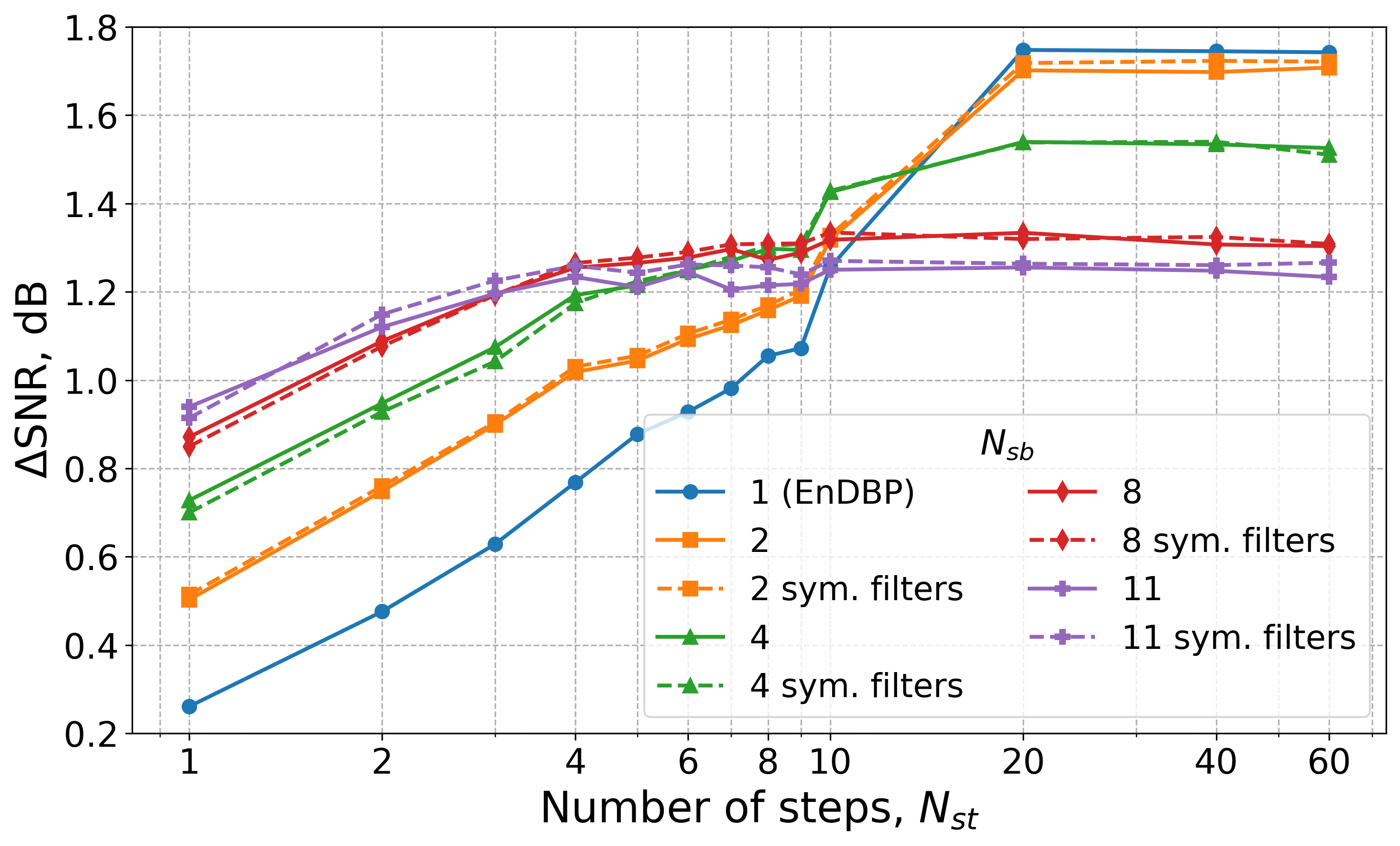}
\caption{SNR gain as a function of the number of steps $N_{st}$ for SbL-DBP with different numbers of subbands $N_{sb}\in\{1,2,4,8,11\}$, the nonlinear memory is fixed to $N_c=8$ (solid curves). Dashed curves correspond to filters optimized with symmetry and translational invariance constraints (sec.~\ref{subsec:mimo_coeff}).}
\label{dSNR_Nst_SbLDBP}
\end{figure}

First, we analyze the impact of subband decomposition on the performance of the proposed method as a function of the number of spatial steps $N_{st}$.
The results are shown in Fig.~\ref{dSNR_Nst_SbLDBP} (solid curves) for different numbers of subbands $N_{sb}\in\{1,2,4,8,11\}$, while the nonlinear memory is fixed to $N_c=8$.
As expected, all curves exhibit a monotonic increase in SNR with increasing $N_{st}$, followed by saturation when further reduction in step size no longer improves the compensation accuracy.
However, a more detailed analysis shows two distinct regimes with different behavior.

In the regime of a small number of steps ($N_{st}<10$), the SNR gain increases with the number of subbands for a fixed $N_{st}$.
In particular, the best performance is achieved for $N_{sb}=11$, while the full-band case ($N_{sb}=1$) provides the lowest gain.
This behavior can be explained by the dominant role of the SSFM approximation error in this regime.
With a limited number of steps, conventional full-band DBP cannot accurately capture the interaction between dispersion and nonlinearity, resulting in significant residual distortions.
Subband decomposition mitigates this limitation by reducing the effective bandwidth within each subband.
As a result, the corresponding dispersion-induced temporal spreading becomes smaller and the nonlinear evolution becomes more localized in time.
This effectively increases the dispersion length within each subband and improves the accuracy of the SSFM approximation.
Consequently, even with a small number of steps, SbL-DBP can more accurately compensate for dominant phase distortions, including SPM within each subband and XPM between subbands. 
Although FWM interactions are not captured in this approach, the improvement in SPM and XPM compensation outweighs the impact of unmodeled FWM.

In contrast, for a sufficiently large number of steps ($N_{st}>10$), the trend reverses: the SNR gain decreases as the number of subbands increases and the best performance is achieved for the full-band case ($N_{sb}=1$).
In this regime, the SSFM approximation error is already small due to the fine discretization of the propagation distance. 
Therefore, performance is no longer limited by numerical accuracy of SSFM, but rather by the ability of the model to capture all relevant nonlinear interactions.
While SPM and XPM are well represented in the subband framework, FWM involves coherent interactions across a broader frequency range and is not fully captured when the signal is decomposed into independently processed subbands with simplified coupling.
The strength and complexity of these inter-subband interactions increase with the number of subbands.
As a result, residual nonlinear distortions accumulate and limit the achievable SNR.

The results highlight a clear trade-off between the number of subbands and the number of spatial steps.
Subband processing is particularly effective in the low-step regime, where it significantly reduces the approximation error and enables accurate nonlinear compensation with limited resources.
However, as the number of steps increases, its relative advantage diminishes and may even become negative due to incomplete modeling of broadband nonlinear interactions, especially FWM.
At the same time, these trends must be interpreted in the context of computational complexity, which plays a critical role in practical system design and is analyzed in the following subsections.

\subsection{Analysis of MIMO Coefficients}
\label{subsec:mimo_coeff}

\begin{figure*}[!b]
\centering
\includegraphics[width=7.1in]{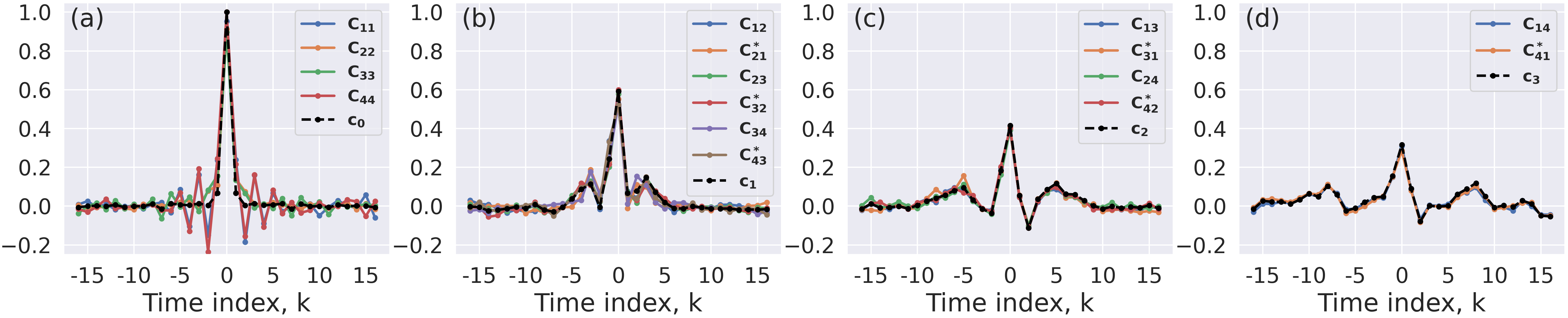}
\caption{Normalized coefficients $C_{jlk}$ for the case $N_{sb} = 4, N_c = 16, N_{st} = 20$ calculated without constraints (colored solid curves) and calculated with constraints on the conditions (\ref{eq:sym_xpm})-(\ref{eq:inv}) (black dashed curves).}
\label{MIMO_C_time}
\end{figure*}

The coefficients of the MIMO filter $C_{jlk}$ form a real-valued three-dimensional tensor of size $N_{sb}\times N_{sb}\times (2N_c + 1)$. 
From perturbation analysis of the NLSE, it follows that these coefficients theoretically satisfy a set of symmetry and translational invariance properties~\cite{civelli_jlt_2025}. 
To better understand the structure of the learned nonlinear interactions, we analyze the learned coefficients both with and without explicitly applying these properties during training.
\begin{enumerate}
    \item The symmetry property can be written as
       \begin{equation}
           C_{jlk} = C_{lj-k}.
           \label{eq:sym_xpm}
       \end{equation}
       In particular, for the coefficients corresponding to SPM, this condition reduces to even symmetry with respect to the time index:
       \begin{equation}
           C_{jjk} = C_{jj-k}.
           \label{eq:sym_spm}
       \end{equation}
   \item The translational invariance property states that the coefficients depend only on the relative distance between the interacting subbands and not on their absolute indices. In other words, they are invariant under a simultaneous shift of the indices:
    \begin{equation}
        C_{jlk} = C_{j+m\,l+m\,k} \quad \forall m: l+m, j+m \in \mathcal{J}.
        \label{eq:inv}
    \end{equation}
\end{enumerate}

Taking into account properties (\ref{eq:sym_xpm}) and (\ref{eq:inv}), the coefficient tensor can be represented by only $N_{sb}$ unique vectors $\mathbf{c}_p$, indexed by the subband separation $p = |j - l|$.
For example, when $N_{sb} = 4$, the full tensor is reduced to four groups corresponding to $p=0,1,2,$ and $3$, i.e., the main diagonal and the three off-diagonal interaction classes:
\[
\mathbf{c}_0: C_{11k}=C_{22k}=C_{33k}=C_{44k},
\]
\[
\mathbf{c}_1: C_{12k}=C_{23k}=C_{34k}=C_{21-k}=C_{32-k}=C_{43-k},
\]
\[
\mathbf{c}_2: C_{13k}=C_{24k}=C_{31-k}=C_{42-k},
\]
\[
\mathbf{c}_3: C_{14k}=C_{41-k}.
\]
Fig.~\ref{MIMO_C_time} shows the learned coefficients for $N_{sb} = 4$, $N_c = 16$, and $N_{st} = 20$. The colored solid curves correspond to the coefficients obtained without imposing the symmetry and translational invariance constraints, while the black dashed curves show the vectors $\mathbf{c}_p$ computed under (\ref{eq:sym_xpm})-(\ref{eq:inv}) constraints. In both cases, the coefficients are normalized with respect to the maximum absolute value. The notation $\mathbf{C}^*_{jl}$ in the figure denotes the time-reversed coefficients, i.e. $C^*_{jlk} = C_{jl-k}$. The corresponding $\Delta SNR$ values are nearly identical for the constrained and unconstrained models, at approximately $1.53$~dB, which confirms that the imposed properties do not degrade performance.

Fig.~\ref{MIMO_C_time}(a) shows the coefficients corresponding to $p = 0$, which describe intra-subband effects. In this case, the central coefficient at $k = 0$, corresponding to SPM, is significantly larger than the remaining coefficients. 
Most significant coefficients are concentrated within the interval $k\in[-8,8]$, while the outer taps remain close to zero. In addition, the vectors $\mathbf{C}_{11}$ and $\mathbf{C}_{44}$, as well as $\mathbf{C}_{22}$ and $\mathbf{C}_{33}$, closely coincide pairwise due to the frequency symmetry of the corresponding subbands.
Although the unconstrained coefficients do not exactly satisfy translational invariance, this has little impact on performance, indicating that multiple coefficient configurations can achieve similar SNR values. For the constrained vector $\mathbf{c}_0$, only the few central coefficients remain significant, while the remaining taps are close to zero.

Figs.~\ref{MIMO_C_time}(b)-(d) show the coefficient vectors for $p = 1$, $2$, and $3$, corresponding to inter-subband XPM interactions. In all cases, the constrained and unconstrained coefficients exhibit very similar structures. Unlike the intra-subband case, the inter-subband coefficients naturally satisfy the symmetry and translational invariance properties even without explicitly enforcing them during training, indicating that their structure is strongly determined by the underlying propagation physics. The largest coefficients are located near $k=0$, and their magnitude decreases with increasing $p$, showing that the interaction strength becomes weaker for more distant subbands. In addition, distinct side peaks appear at $k=\pm3$ for $p=1$, $k=\pm5$ for $p=2$, and $k=\pm8$ for $p=3$. These peaks correspond to the relative group delay between subbands induced by chromatic dispersion. As the frequency separation between subbands increases, the relative walk-off also increases, shifting the interaction maxima toward larger values of $|k|$.

Beyond the coefficient-level analysis, the impact of the imposed constraints on the overall system performance is further validated in Fig.~\ref{dSNR_Nst_SbLDBP}. The solid curves correspond to unconstrained MIMO filters, while the dashed curves represent models trained with the symmetry and translational invariance constraints applied (labeled ``sym. filters'' in the legend). As can be seen, the constrained models closely follow the performance of their unconstrained counterparts across all considered configurations, with only minor deviations caused by optimization variability. These results confirm that enforcing properties (\ref{eq:sym_xpm})-(\ref{eq:inv}) preserves the nonlinear compensation performance, while significantly reducing the number of independent coefficients.

The symmetry and translational invariance properties (\ref{eq:sym_xpm})-(\ref{eq:inv}) provide a reduction in computational complexity, specifically in the weighted summation with MIMO coefficients at the nonlinear step. By exploiting these properties, the number of unique coefficients is reduced from $N_{sb}^2 (2N_c + 1)$ to $(N_{sb}-1) (2N_c + 1) + (N_c + 1)$, while redundant computations between equivalent coefficient groups are eliminated. As a result, the computational complexity of the nonlinear step per processed symbol becomes
\begin{equation}
\begin{aligned}
C^{sym}_{nl} 
&= N_{st} \frac{\frac{3}{2}N \left(N_{sb}\left(N_c + \frac{1}{2}\right) + 5\right)}{(R_s/F_s)N - N_d}.
\end{aligned}
\label{eq:cnln_MIMO_sym}
\end{equation}
All subsequent results for the SbL-DBP method are obtained using the constrained formulation, which provides improved computational efficiency without a noticeable loss in SNR performance.

\subsection{Sparsification of MIMO Coefficients}
\label{subsec:mimo-coeff}
To further reduce the computational complexity of SbL-DBP, we investigate pruning of the learned MIMO filters. The main idea is to remove coefficients that contribute only marginally to nonlinear compensation while preserving the dominant nonlinear interactions responsible for SNR gain. To achieve this, an additional regularization term is incorporated into the loss function according to (\ref{eq:reg}). This regularization penalizes a large number of nonzero coefficients and encourages low-magnitude coefficients to approach zero during training.

\begin{figure}[!htbp]
\includegraphics[width=3.4in]{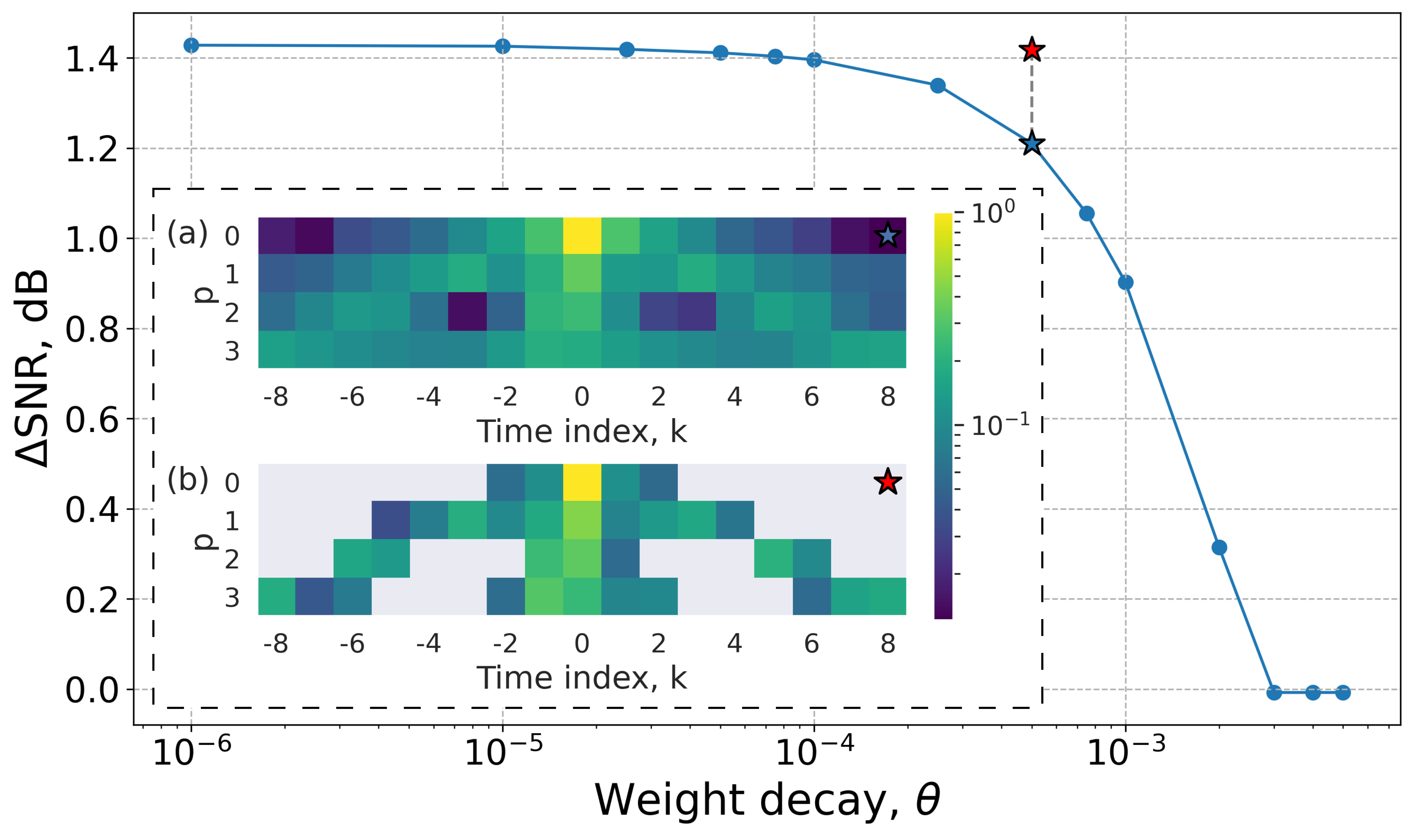}
\caption{SNR gain $\Delta SNR$ as a function of the weight decay $\theta$ for the SbL-DBP with $N_{st} = 10$, $N_{sb} = 4$, and $N_c = 8$. Inset: (a) distribution of the learned MIMO coefficient magnitudes for $\theta = 5 \cdot 10^{-4}$, (b) coefficient distribution after sparsification.}
\label{dSNR_decay}
\end{figure}

To determine the optimal regularization strength, we analyze the dependence of the achieved $\Delta SNR$ on the regularization weight $\theta$. The corresponding results are shown in Fig.~\ref{dSNR_decay} for the SbL-DBP configuration with $N_{st} = 10$, $N_{sb} = 4$, and $N_c = 8$. Without regularization ($\theta=0$), the model achieves $\Delta SNR=1.43$~dB. As the regularization weight increases, the magnitude of many coefficients decreases, leading to a progressively sparser nonlinear operator. At the same time, strong regularization degrades compensation accuracy and reduces the achieved SNR gain.

Based on this trade-off, we select the operating point corresponding to $\theta = 5 \cdot 10^{-4}$, indicated by the blue star marker in Fig.~\ref{dSNR_decay}. 
At this point, the model still preserves most of its nonlinear compensation capability while already exhibiting a pronounced concentration of low-magnitude coefficients. The corresponding distribution of the learned coefficient magnitudes is shown in the inset~(a) of Fig.~\ref{dSNR_decay}.

Guided by this distribution, we apply an additional pruning stage by setting to zero all coefficients with magnitudes below a threshold of $4.5 \cdot 10^{-2}$. 
This threshold is chosen to balance sparsity and performance degradation.
As a result, 35 out of 68 coefficients (51.5\%) are removed from the nonlinear operator. 
After pruning, the remaining nonzero coefficients are further fine-tuned through additional training without regularization in order to compensate for the performance loss caused by sparsification.
The resulting coefficient distribution is shown in inset~(b) of Fig.~\ref{dSNR_decay}.

Importantly, despite the more than twofold reduction in the number of active coefficients, the performance degradation remains negligible.
After retraining, the pruned model achieves $\Delta SNR=1.42$~dB (red star marker), compared to $1.43$~dB for the original dense model.
These results indicate that the SbL-DBP framework contains a substantial degree of parameter redundancy and that sparsification can significantly reduce computational complexity while preserving nearly the same nonlinear compensation performance.

\subsection{Performance--Complexity Trade-off}
\begin{figure}[!htbp]
\includegraphics[width=3.4in]{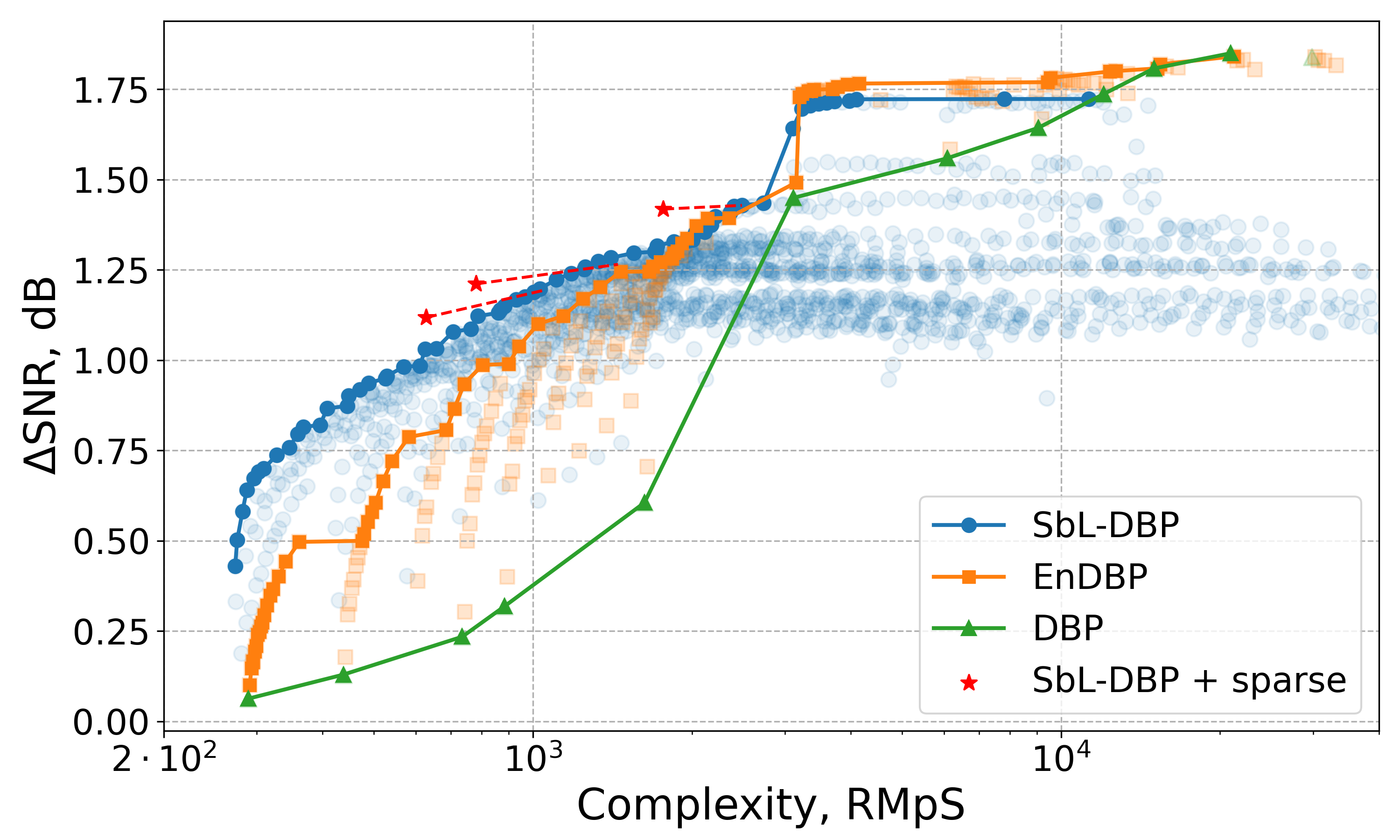}
\caption{Improvement of the signal-to-noise ratio $\Delta SNR$ as a function of the computational complexity for DBP-based methods: DBP (green triangles), EnDBP (orange squares indicating all realizations for different $N_c$ and $N_{st}$ values), and SbL-DBP (blue circles indicating all realizations for different $N_{sb}$, $N_{c}$, and $N_{st}$ values). The curves represent the Pareto-optimal sets. Red star markers represents SbL-DBP with sparsified MIMO coefficients.}
\label{dSNR_Cmplxt}
\end{figure}

Next, we evaluate the performance of the SbL-DBP method in terms of the trade-off between SNR gain and computational complexity and compare it with DBP and EnDBP.
To ensure a fair and comprehensive comparison, we consider the Pareto-optimal performance of each method.
For this, a wide range of parameter configurations is evaluated for each approach, and the Pareto front is then constructed by selecting only those configurations for which no other configuration achieves both higher SNR and lower complexity.
For DBP, the optimization is performed over the number of spatial steps $N_{st}$, for EnDBP -- over $N_{st}$ and nonlinear memory $N_c$, for SbL-DBP -- over $N_{st}$, $N_c$ and the number of subbands $N_{sb}$.
Fig.~\ref{dSNR_Cmplxt} shows the SNR gain as a function of complexity, where all evaluated configurations are depicted as transparent markers and the resulting Pareto fronts for all methods as solid curves.

It can be seen that, across the entire range of computational complexity, DBP consistently provides the lowest performance.
EnDBP significantly improves upon DBP by incorporating a nonlinear step with memory, which partially captures the interaction between dispersion and nonlinearity.
As a result, EnDBP achieves a substantially better performance--complexity trade-off, particularly in the low- and medium-complexity regimes. In the high-complexity regime the performance of DBP and EnDBP becomes comparable.
Most importantly, the proposed SbL-DBP method demonstrates a clear advantage over both DBP and EnDBP in the low- and medium-complexity regimes.
For example, at a complexity of approximately 500~RMpS, SbL-DBP ($N_{sb}=13$, $N_{st}=1$, $N_c=6$) achieves an SNR gain of about 0.94~dB, compared to 0.58~dB for EnDBP ($N_{st}=2$, $N_c=14$) and 0.13~dB for DBP ($N_{st}=2$).
Similarly, at a complexity of approximately 1600~RMpS, SbL-DBP ($N_{sb}=6$, $N_{st}=4$, $N_c=12$) achieves an SNR gain of about 1.3~dB, compared to 1.24~dB for EnDBP ($N_{st}=8$, $N_c=20$) and 0.61~dB for DBP ($N_{st}=10$).

The observed performance advantage can be interpreted as follows.
By decomposing the signal into subbands, SbL-DBP reduces the bandwidth and channel memory within each subband, enabling more accurate modeling of nonlinear phase evolution.
At the same time, inter-subband interactions are captured through a learnable MIMO structure.
As a result, the increase in computational complexity grows more slowly than the corresponding improvement in SNR, leading to a more favorable performance--complexity trade-off.

In the high-complexity regime (above $\sim 4\cdot10^3$ RMpS), the gap between SbL-DBP and EnDBP becomes smaller, and EnDBP may slightly outperform SbL-DBP for certain configurations. 
This behavior is consistent with the observations in Section~\ref{subsec:subband_and_num_steps}. 
In this regime, the main limitation is the ability to capture broadband nonlinear interactions, such as FWM, which are only partially modeled in the subband framework. 

In addition, Fig.~\ref{dSNR_Cmplxt} includes several SbL-DBP configurations with sparsified MIMO coefficients, indicated by red star markers. These points demonstrate that applying coefficient pruning after training can further improve the performance--complexity trade-off.

\subsection{Optimal Number of Subbands}
\begin{figure}[!htbp]
\includegraphics[width=3.4in]{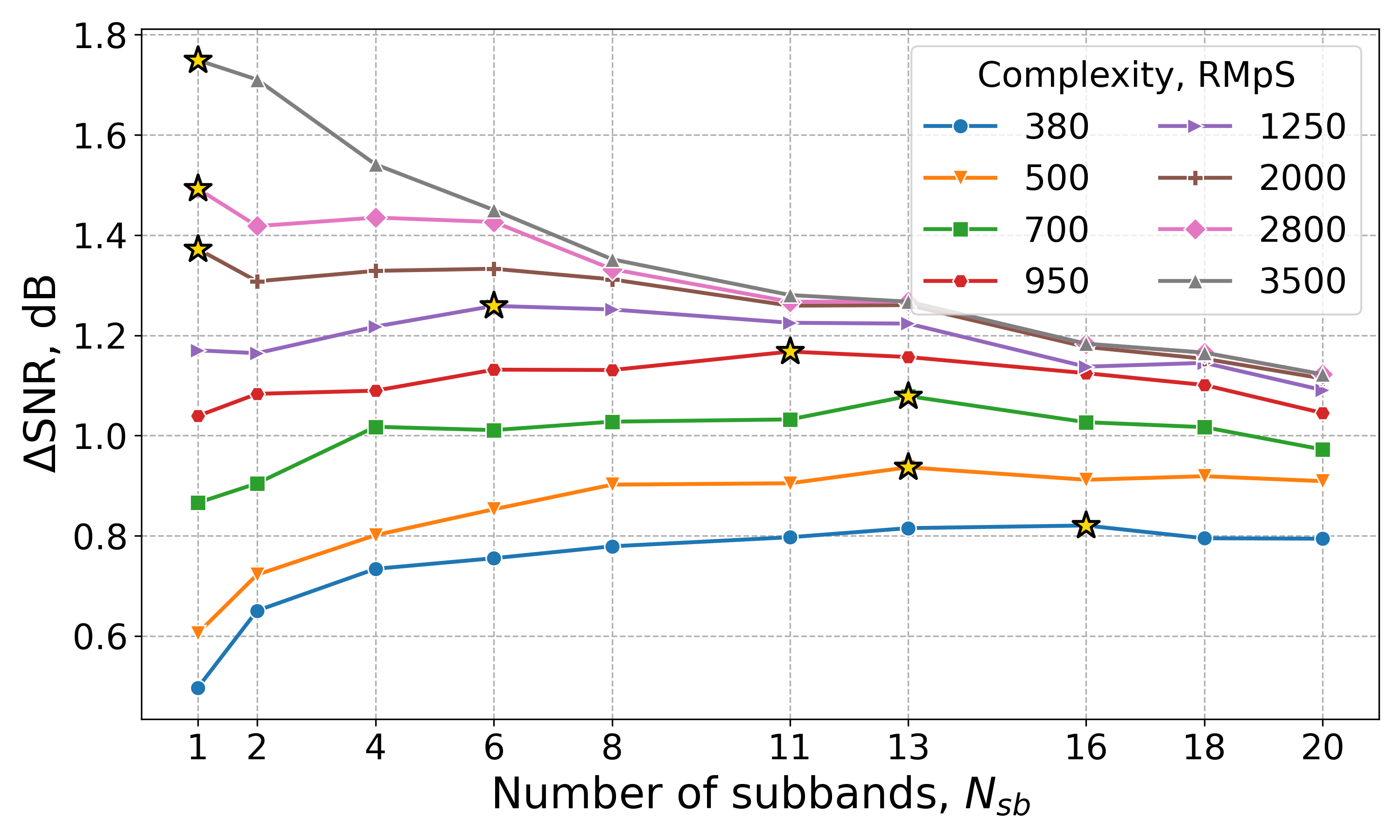}
\caption{SNR gain as a function of the number of subbands $N_{sb}$ for SbL-DBP under different computational complexity constraints.}
\label{dSNR_Nsb}
\end{figure}

Here we investigate the selection of the optimal number of subbands $N_{sb}$ for the proposed SbL-DBP method. 
Fig.~\ref{dSNR_Nsb} shows the SNR improvement achieved by SbL-DBP as a function of the number of subbands for several complexity levels. The markers correspond to different SbL-DBP configurations with fixed $N_{sb}$, while the parameters $N_{st}$ and $N_c$ are adjusted to satisfy a predefined complexity budget. The star markers indicate the operating points that provide the maximum $\Delta\text{SNR}$ for each complexity level.

The obtained results show that the optimal number of subbands strongly depends on the target computational complexity.
In the low-complexity regime, larger values of $N_{sb}$ are preferable, with the best performance achieved for approximately $N_{\mathrm{sb}}=16$. 
In this regime, the dominant limitation is the SSFM approximation error caused by the small number of spatial steps. Subband decomposition reduces the effective bandwidth and channel memory within each subband, which improves the accuracy of nonlinear compensation and enables efficient processing even with a limited number of steps. 
However, further increasing the number of subbands beyond the optimal value leads to performance degradation due to the increasing importance of broadband nonlinear interactions, particularly FWM, which are not fully captured within the considered subband framework.

In contrast, in the high-complexity regime, lower values of $N_{sb}$ become optimal. In this case, the SSFM approximation error is already sufficiently small due to the larger number of spatial steps, and the main limitation shifts to the accuracy of nonlinear interaction modeling. Using fewer subbands better preserves broadband nonlinear interactions while still benefiting from moderate channel memory reduction. Therefore, the optimal choice of $N_{\mathrm{sb}}$ represents a trade-off between improved numerical accuracy obtained through subband decomposition and the ability to accurately model broadband nonlinear effects.

\subsection{Performance versus Launch Power}
\begin{figure}[!htbp]
\includegraphics[width=3.4in]{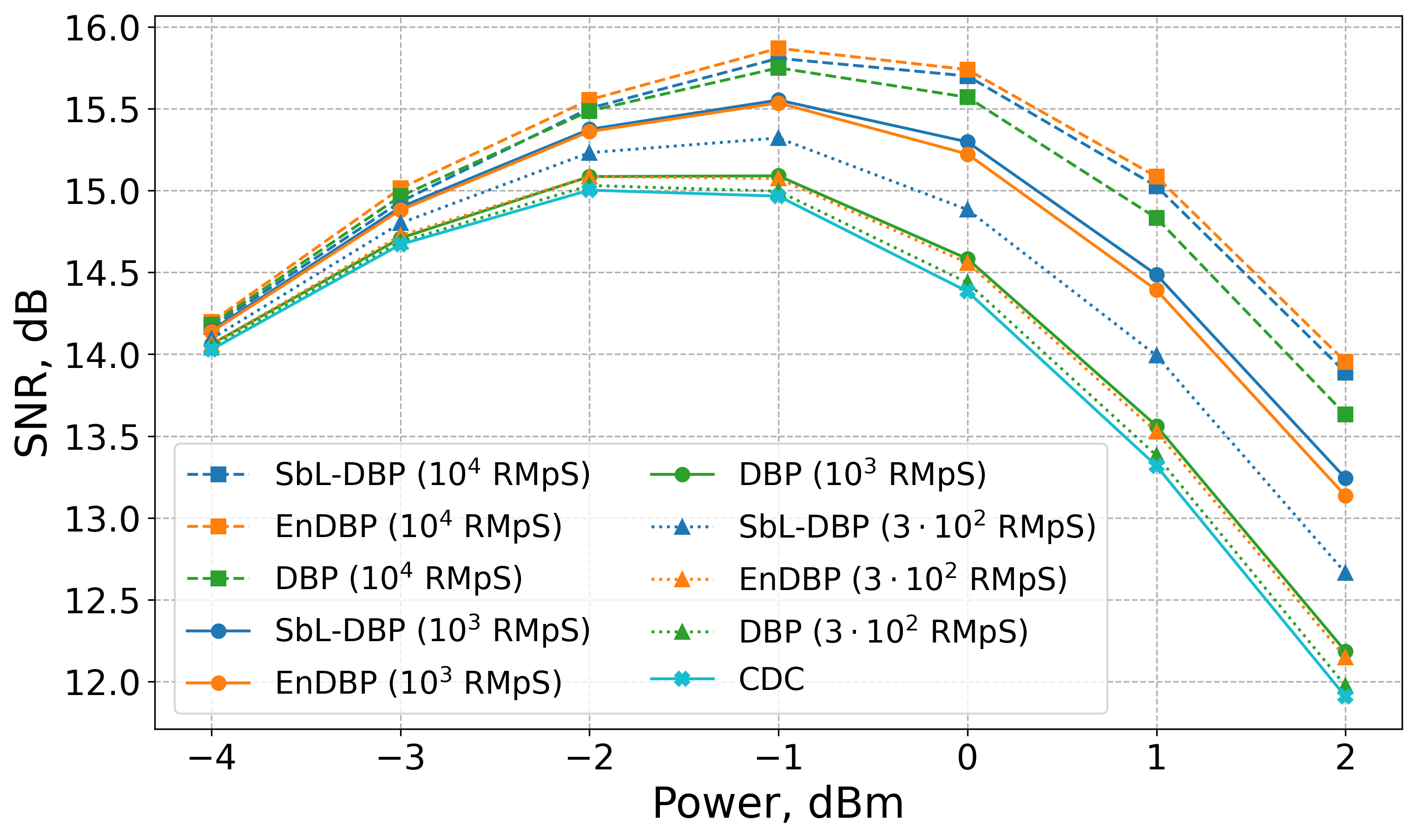}
\caption{SNR as a function of launch power per channel for SbL-DBP, EnDBP, and DBP under equal computational complexity constraints. Three representative complexity regimes are considered: low ($\approx 3\cdot10^2$ RMpS, dotted), medium ($\approx 10^3$~RMpS, solid), and high ($\approx 10^4$~RMpS, dashed).}
\label{SNR_powers}
\end{figure}

To further evaluate the robustness of the proposed method, we compare the performance of SbL-DBP with EnDBP and DBP as a function of the launch power in the range from -4 to 2~dBm.
To ensure a fair comparison, all methods are evaluated under equal computational complexity constraints. Three representative complexity levels are considered: (i) low complexity ($\approx 3\cdot10^2$~RMpS), (ii) medium complexity ($\approx 10^3$~RMpS), and (iii) high complexity ($\approx 10^4$~RMpS). 
For each complexity level, the parameters of all methods are selected to maximize the achieved SNR while ensuring that the total computational cost matches the target value. The resulting dependence of the achieved SNR on the launch power per channel is shown in Fig.~\ref{SNR_powers}.

For the low-complexity regime, the following configurations are used: $N_{st}=1$, $N_c=1$, $N_{sb}=16$ (SbL-DBP); $N_{st}=1$, $N_c=5$ (EnDBP); $N_{st}=1$ (DBP).
In this regime, all methods operate with a very limited number of steps, which makes the approximation error of nonlinear compensation dominant.
Across the entire power range, SbL-DBP consistently provides the highest SNR.
The gain is moderate at low launch powers (noise-limited regime), but becomes more pronounced as the power increases.
At 1~dBm, which corresponds to the operating point used in the previous sections, SbL-DBP provides an SNR gain of approximately 0.47~dB over EnDBP and 0.61~dB over DBP.

For the medium-complexity regime, the following configurations are used: $N_{st}=3$, $N_c=4$, $N_{sb}=11$ (SbL-DBP); $N_{st}=4$, $N_c=30$ (EnDBP); $N_{st}=6$ (DBP).
At this complexity level, all methods achieve improved performance due to increased model completeness. 
However, the relative trends remain similar.
SbL-DBP continues to outperform both EnDBP and DBP across all launch powers, despite using a smaller number of steps.
The performance gain is especially visible in the nonlinear regime (around 0 to 2~dBm), where nonlinear distortions dominate.
Compared to EnDBP, SbL-DBP achieves an SNR improvement of approximately 0.07-0.11~dB, while the gap to DBP remains larger.

For the high-complexity regime, the following configurations are used: $N_{st}=40$, $N_c=16$, $N_{sb}=2$ (SbL-DBP); $N_{st}=60$, $N_c=8$ (EnDBP); $N_{st}=65$ (DBP).
At this complexity, all methods approach their performance limits and the differences between them become smaller.
This reduced gap compared to lower complexity regimes is expected, as increasing the number of steps improves the accuracy of all methods.
However, in this regime, EnDBP consistently outperforms SbL-DBP across the considered power range, likely due to its more effective modeling of FWM interactions.

\section{Conclusion}
We have proposed a subband-based learned digital backpropagation framework for compensation of nonlinear distortions in wideband optical communication systems. The proposed approach combines subband signal decomposition with a trainable DBP structure, where chromatic dispersion is compensated independently in each subband in the frequency domain, while nonlinear intra- and inter-subband interactions are modeled in the time domain using a learnable MIMO filtering structure. All model parameters are jointly optimized using end-to-end gradient-based learning.

By reducing the bandwidth of individual subbands, the proposed framework decreases the effective channel memory and relaxes the requirements on the spatial discretization step, enabling more computationally efficient modeling of nonlinear interactions. In addition, the proposed sparsification procedure reduces the number of active nonlinear coefficients by more than a factor of two while preserving nearly the same compensation performance.

Numerical simulations of a wideband 11$\times$40~Gbaud WDM RRC-16QAM 20$\times$100 km transmission system demonstrate that the proposed SbL-DBP framework provides a superior performance--complexity trade-off compared with conventional DBP and EnDBP methods, particularly in the low- and medium-complexity regimes. In these regimes, the approach achieves higher SNR gains while requiring fewer propagation steps. Furthermore, the method demonstrates stable performance over a wide range of launch powers.

The obtained results indicate that combining subband processing with a learnable modeling of nonlinear interaction is a promising direction for low-complexity nonlinear compensation in future high-capacity wideband optical communication systems.

\section*{Acknowledgments}
This research was funded by the Russian Science Foundation (Grant No. 25-61-00010), https://rscf.ru/project/25-61-00010/.

\vfill

\end{document}